\documentclass[fleqn,usenatbib]{mnras}

\usepackage{newtxtext,newtxmath}

\usepackage[T1]{fontenc}
\usepackage{ae,aecompl}
\usepackage{threeparttable}
\usepackage{blindtext}

\usepackage{graphicx}	
\usepackage{amsmath}	

\title[Metallicity variation in NGC 1261]{ Intrinsic Metallicity Variation  in  the Intermediate Mass Type II Globular Cluster NGC~1261}

\author[C. Mu\~noz et al.]{
C. Mu\~noz$^{1,2,3}$\thanks{E-mail:cesar.alejandro.munoz.g@gmail.com},
D. Geisler$^{1,2,3}$,
S. Villanova$^{3}$,
Ata Sarajedini$^{4}$,
H. Frelijj$^{3}$,
C. Vargas$^{3}$,
\newauthor
L. Monaco$^{5}$,
J. O\textquotesingle Connell$^{3}$
\\
\\
$^{1}$Instituto de Investigaci\'on Multidisciplinario en Ciencia y Tecnolog\'ia, 
Universidad de La Serena. Avenida Ra\'ul Bitr\'an S/N, La Serena, Chile.\\
$^{2}$Departamento de  Astronom\'ia, Facultad de Ciencias, Universidad de La Serena. Av. Juan Cisternas 1200, La Serena, Chile.\\
$^{3}$Departamento de Astronom\'ia, Casilla 160-C, Universidad de
  Concepci\'on, Concepci\'on, Chile.\\
$^{4}$Department of Physics, Florida Atlantic University, 777 Glades Rd, Boca Raton, FL 33431, USA.\\
$^{5}$Departamento de Ciencias Fisicas, Universidad Andres Bello, Fernandez Concha 700, Las Condes, Santiago, Chile
}

\date{Accepted XXX. Received YYY; in original form ZZZ}

\pubyear{2015}

\begin{document}
\label{firstpage}
\pagerange{\pageref{firstpage}--\pageref{lastpage}}
\maketitle

\begin{abstract}

Globular Clusters (GCs)  are now well known to almost universally show multiple populations (MPs). The HST UV Legacy Survey of a large number of Galactic GCs in UV filters optimized to explore  MPs finds that a small fraction of GCs, termed Type II, also display more complex, anomalous behavior. Several well-studied Type II GCs show intrinsic Fe abundance variations, suggesting  that the other, less well-studied, Type II GCs should also exhibit similar behavior.

Our aim is to perform the first detailed metallicity analysis of  NGC~1261, an intermediate mass Type II GC, in order to determine if this object shows an intrinsic Fe variation.  We determined the Fe  abundance in eight  red giant  members using Magellan-MIKE and UVES-FLAMES high-resolution, high S/N spectroscopy. 

The full range of [Fe/H] for the entire sample from the spectra is from -1.05 to -1.43 dex with an observed spread $\sigma_{obs}$=0.133 dex. Compared with the total internal error of  $\sigma_{tot}$=0.06, this indicates a significant intrinsic metallicity spread of $\sigma_{int}$=0.119 dex. We found a very similar variation in [Fe/H] using an independent method  to derive the atmospheric parameters based on  near-IR photometry. More importantly, the mean metallicity of the five  presumed normal metallicity stars  is -1.37$\pm 0.02$, while that of the three presumed anomalous/high metallicity stars is -1.18$\pm0.09$. This difference  is significant at the $\sim2.4\sigma$ level.

  We find indications from existing data of other Type II GCs that  several  of them presumed to have real metallicity spreads may in fact posses none. The minimum mass required for a GC to acquire an intrinsic Fe spread appears to be $\sim10^5 M_\odot$. We find no strong correlation between mass and metallicity variation for Type II GCs. The metallicity spread  is also independent of the fraction of anomalous stars within the Type II GCs and of GC origin.

\end{abstract}


%
%
%
\section{Introduction}

The more closely we examine globular clusters (GCs), the more complex and interesting we discover them to be. Long regarded as quintessential Simple Stellar Populations, we now know them to instead host multiple populations (MPs), showing characteristic intracluster chemical variations involving principally light elements, most notably C, N, O, Na, Mg and/or Al \citep{carretta09a}. This of course has caused a paradigm shift in our view of GC formation, and we have had to discard the traditional wisdom that they are made entirely of coeval and (initially) chemically homogeneous stars, all formed during a single star formation burst. 

Following the current most widely accepted interpretation, observational
evidence strongly suggests that multiple generations of stars formed within a GC over a relatively short timescale ($\lesssim$100 Myr) from gas polluted by the products of hot proton-capture processes which occurred in the interiors of the initial generation of intermediate mass asymptotic giant branch stars, or massive fast rotating/binary
stars, which subsequently reached the stellar surface and then escaped via stellar winds but was retained by the GC potential well to form a later generation(s)
\citep[e.g.][]{Dercole08,decressin07,Demink09}. Note that, however, none of the various scenarios currently suggested is able to completely reproduce the vast array of complex behavior \citep{renzini15}.

On the other hand, homogeneity in the cluster iron abundance is a much more common phenomenon. This is generally attributed to the likelihood that, although typical GCs were massive enough to retain the ejecta from low velocity stellar winds responsible for the light element pollution manifested by MPs, they were not massive enough to retain the ejecta of much more energetic supernovae (SNe) responsible for heavy element production, unlike even dwarf galaxies. There are however a few notable exceptions to this general rule, such as $\omega$ Centauri, M54 and Terzan 5 \citep{meszaros20,Massari14b,villanova10,villanova14,johnso10,pancino02,marino11a,ferraro09}. These extreme, very massive objects are now generally regarded as the remnants of accreted  nucleated dwarf galaxies (M54 is the central cluster in the Sgr dwarf galaxy \citep{bellazzini08}) and of a basic building block of the Galactic bulge (Ferraro et al. 2020), respectively. The nature of the very limited number of other, lower mass GCs where smaller ($\sim$0.1 dex) Fe spreads have been measured, such as M22 \citep{Marino09}, is less certain, as indeed is the reality of an intrinsic metallicity spread in certain of these GCs \citep[e.g.][]{mucciarelli15}. The origin of these anomalous clusters is highly debated, with the most favored possibilities being that they are remnants of disrupted dwarf galaxies, or above the mass threshold required to retain SNe ejecta, or perhaps the products of GC mergers.

The largest and most homogeneous as well as most detailed compilation of photometric MP behavior in Galactic GCs is found in the series of papers using HST WFC3 observations in the so-called “magic trio” of UV filters, namely F275W, F336W and F438W, which were chosen for their high sensitivity to CN, CH, NH and OH molecular band strengths \citep{Piotto15}.   In this UV Legacy Survey of Galactic Globular Clusters, they find a truly bewildering variety of MP behavior: the 57 GCs they observe show 57 different color distributions. However, they \citep{Milone17} are able to identify certain generic patterns and define two classes of GCs based on their distribution in a two-color diagram formed using the magic trio, termed the “chromosome map”. Type I clusters show the typical bimodality where the generic first (1G) and second (2G) stellar generations are almost always clearly separated and no other strong pattern emerges. But $\sim$20$\%$ of the GCs in the sample  studied by \citet{Milone17} are labelled Type II, which  display more complex chromosome maps, with 1G and/or 2G sequences that appear to be split, with an additional distribution of anomalous red stars in this diagram.

By using spectroscopic data from the literature, they showed that Type I GCs only show light element variations, while 
the Type II clusters so far well studied host populations that are also enriched in overall CNO abundance (C+N+O) and heavy elements, such as iron and/or s-process elements. Type II clusters also exhibit multiple SGBs in  the ultraviolet and  optical color-magnitude diagrams, with the fainter SGB joining onto a red RGB, while type I clusters only  show these  features  in the ultraviolet color-magnitude diagrams.

These sequences are populated by stars with enhanced heavy-element abundance, particularly in iron \citep{Milone17}. They argue that 1) split 1G and 2G sequences in the chromosome maps, 2) split SGBs and 3) non uniformity of the iron and s-element abundances, must all be physically connected to each other. This evidence indicates that the chromosome map is an efficient tool to identify candidate GCs with internal variations of heavy elements, which are, again, quite rare and of special importance for understanding GC formation and evolution. All the known clusters with a significant iron spread like $\omega$ Cen and M54 belong to Type II (Terzan 5 was not studied by the UV Legacy team due to its heavy reddening).
If we currently have only a very rough idea of how Type I clusters form, we have much less of an idea how Type II clusters were born.

Interestingly, their classification includes several Type II GCs that were not known to have Fe abundance spreads but lacked detailed high-resolution study and are therefore of prime importance for an in-depth chemical investigation.  Several of these clusters are relatively accessible but only poorly studied in the past with respect to any possible spread in the heavy elements, and neither shows any strong indication for such a spread. \citet{Milone17}
estimate that these clusters could host up to 10\% of stars having an iron content larger than the main population. According to their analysis, the iron abundance spread could be of the order of 0.1-0.2 dex, so easily measurable with sufficient quality high resolution spectra of a large enough sample of the appropriate stars.

Given the importance of GCs with metallicity variations and our lack of understanding of their origin and any possible link between metallicity spreads  and generic MP light-element variations, it is of great interest to investigate any potential GC with such variations with high-resolution spectroscopy. Of particular interest are the lowest mass Type II GCs, as these would have the most difficult time retaining supernova ejecta in order to establish metallicity variations, assuming their initial mass is similar to that of the present-day GC. The two lowest mass Type II GCs are NGC 1261 and NGC 6934. More recently, \citet{marino18} observed four stars at high resolution in NGC 6934, the lowest mass Type II GC, and indeed discovered evidence for an intrinsic metallicity spread, making the detailed study of the other low mass Type II GC even more imperative. In the case of NGC 1261  \citet{filler12} presented the analysis of three stars using high resolution spectra. They found  [Fe/H]=-1.19 $\pm$0.02 ($\sigma=0.01$). Unfortunately this study is only an abstract, and a detailed analysis has not been presented subsequently, not to mention the small sample.

In this article, we investigate the next lowest mass Type II GC, NGC~1261, via high-resolution spectroscopy of eight  red giant branch (RGB) members in order to search for the putative intrinsic iron spread,
We have analyzed these spectra in detail to obtain the stellar parameters and metallicity. A future paper will report on abundances of a large number of other species. 

In section 2 we describe the target selection, in section 3 we describe  the observations and data reduction and in section 4  we explain the methodology we used to calculate  atmospheric parameters and the iron abundance. In section 4 we present our main result concerning any metallicity variation, finding evidence for a real spread. Finally in Section 6 we discuss this result and its implications and present our conclusions.

\section{Target Selection}

\begin{figure}
\centering

  \includegraphics[width=3.6in,height=2.7in]{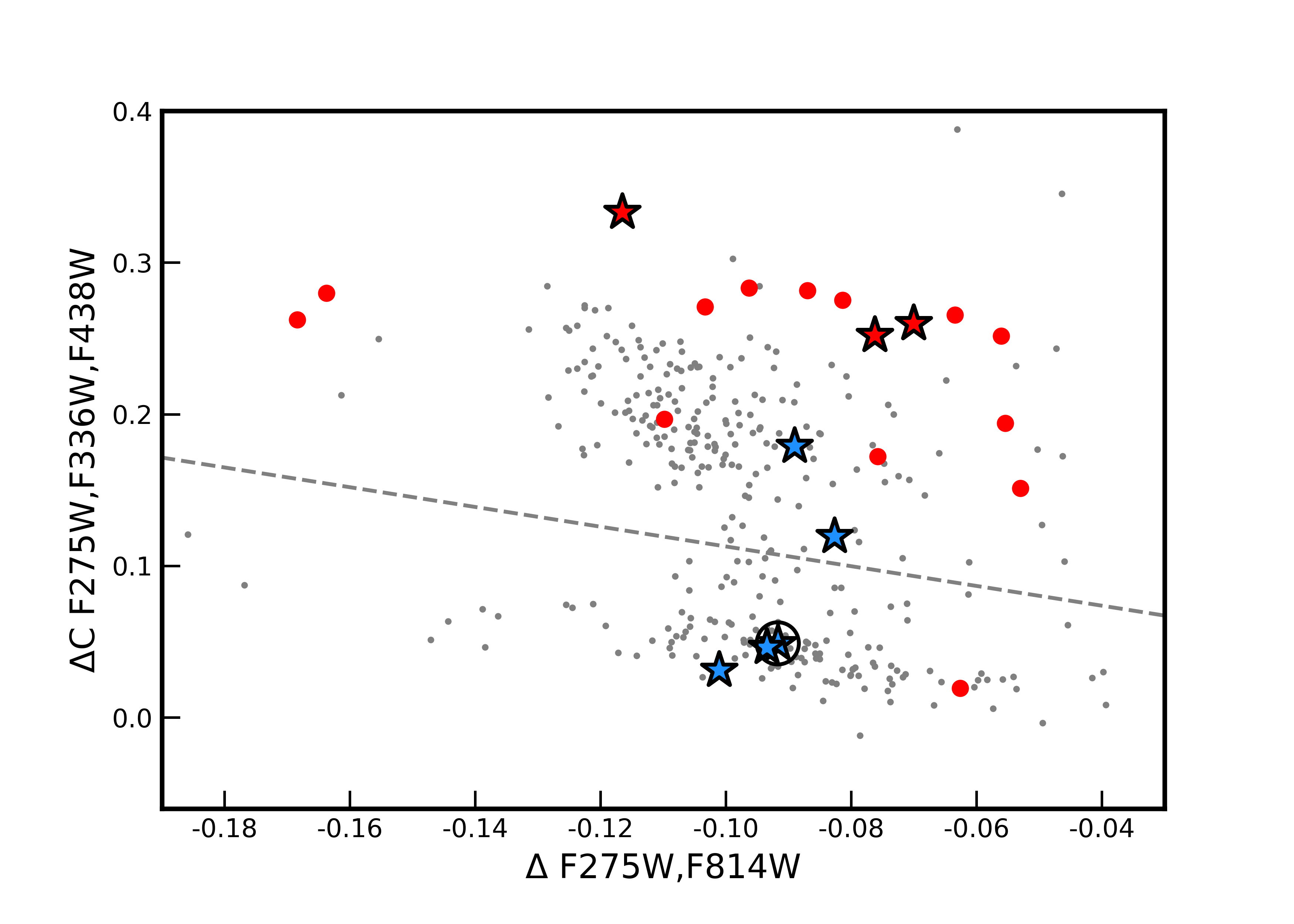}

  \caption{Chromosome map of  NGC~1261.  The red points are the redder RGB stars on the CMD of \citet{Milone17}, as plotted in Figure \ref{CMD}. The dashed line separates 1G (below line) from 2G (above line) stars  according to their definition. Our targets appear as red and blue filled star symbols. The blue targets are assumed to be normal metallicity (in this case 1G stars) and the red targets anomalous, presumably higher metallicity, (in this case 2G) stars, according to the scenario of \citet{Milone17}. The blue target enclosed by a circle is star  \#3,  with data from both MIKE and UVES (See Table \ref{param2}). }
  \label{Chrm}
 \end{figure}

\begin{figure*}
\centering

  \includegraphics[width=5.4in,height=4.0in]{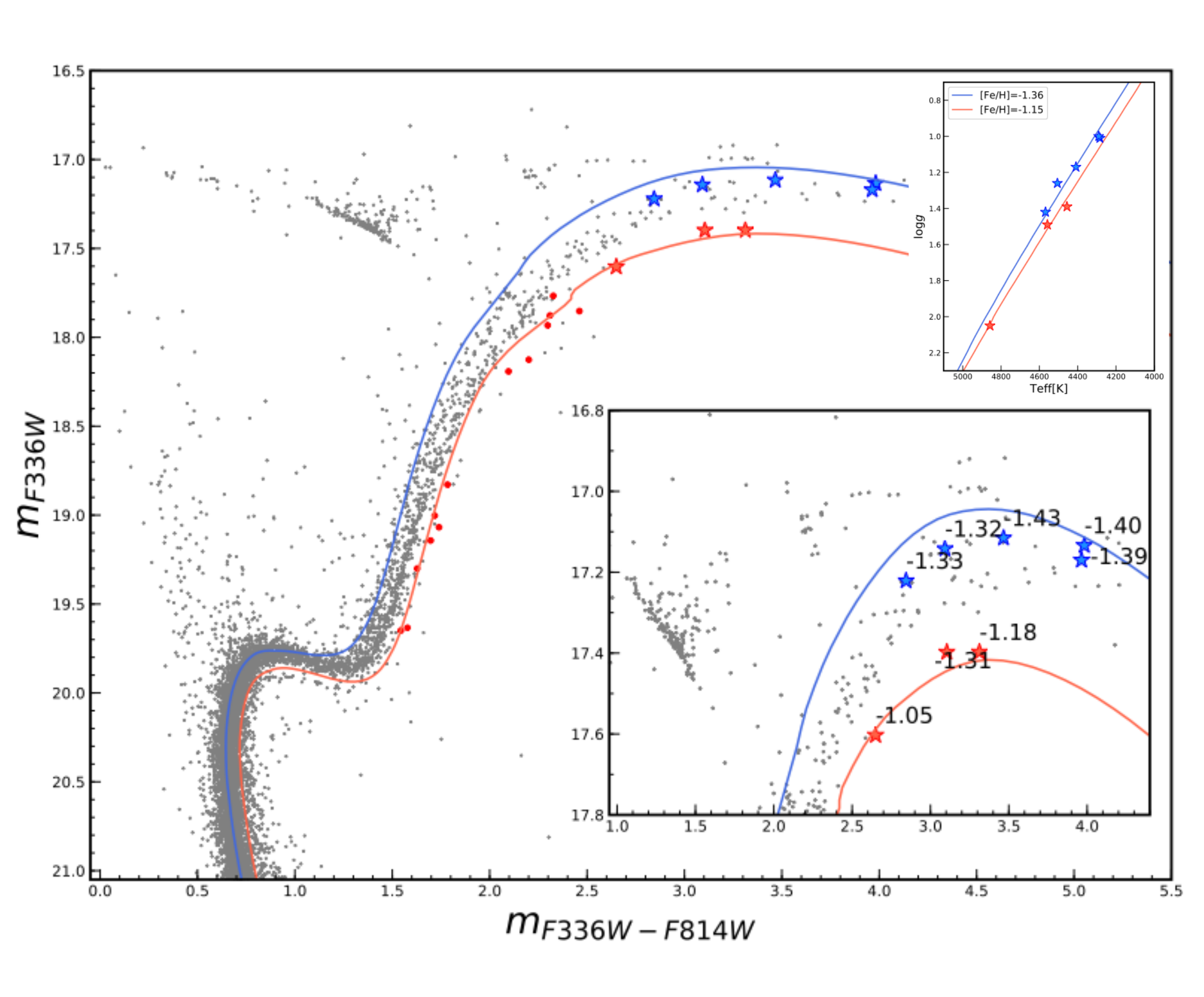}

  \caption{mF336W vs. (mF336W–mF814W) CMD for NGC~1261. The blue and red filled stars represent our spectroscopic targets. The red and blue curves represent theoretical isochrones with an age of 10.75 Gyr \citep{vandenberg13}, [$\alpha$/Fe]=+0.40 dex and a metallicity of -1.15 dex and -1.36 dex respectively  \citep[Dartmouth isochrone]{dotter08}. The lower right insert is a zoom of the upper RGB and includes the metallicity we derive for our targets. The upper right insert shows log$g$ vs. Teff for our targets  measured  spectroscopically and the lines represent the same isochrones.}
  \label{CMD}
 \end{figure*}

We first reproduced the chromosome map of NGC 1261.  To do this, we used the F275W, F336W, F438W and F814W magnitudes from the HST Legacy Survey,  acquired through private communication with A. Sarajedini \citep{Piotto15}, 
and formed the two pseudo-colors $\Delta$F275W,F814W and $\Delta$CF275W,F336W,F438W. The chromosome map is plotted in Figure \ref{Chrm} and clearly shows the dominant presence of typical 1G and 2G stars (black points), relatively well separated. We also plotted the F336W vs. (F336-F814W) CMD (see Figure \ref{CMD}). This plot shows two and perhaps three separate RGBs and a main as well as a loosely populated fainter SGB. Following the method presented in \citet{Milone17}, we plot in red stars in this CMD which lie along the reddest RGB (which merges with the faint SGB), and mark these same stars as red in the chromosome map (Figure \ref{Chrm}). These stars generally fall to redder $\Delta$F275W,F814W colors of the main locus of 2G stars. These outliers (hereby anomalous stars) were identified by \citet{Milone17} and led to their classification of NGC 1261 as a Type II GC. These anomalous stars presumably have a higher metallicity than the bulk of the stars that fall along the nominal 1G and 2G sequences (hereby normal stars), according to their scenario. Thus, we need to observe as many of these anomalous stars as possible, along with a similar number of normal (metallicity)  1G/2G stars, in order to optimize the possibility of detecting any putative metallicity spread between them \citep{marino18}. Note that \citet{Milone17}
finds only about 4$\%$ of RGB stars in NGC 1261 are anomalous, so that careful selection to include a few of these stars is critical. Our final targets included the brightest  anomalous stars from the chromosome map along with bright normal stars chosen from the CMD with (F336W-F814W) colors similar to those of the selected anomalous stars but which lay along the bluer/brighter RGB, in order to maximize the possibility to detect any variation in iron spread. 

\section{Observations and data reduction}
We observed six  RGB  stars in NGC~1261 with the Magellan Inamori Kyocera Echelle (MIKE) spectrograph \citep{Bernstein03} on the 6.5m Clay telescope at the Las Campanas Observatory (LCO) in November, 2018  (CNTAC program ID CN2018B-71, PI D. Geisler ). We observed 3 pairs of anomalous and normal stars, using the  red arm of the spectrograph, which covers a wide wavelength range from 4900-9500\AA,  with a resolving power of  R$\sim$42000 with an 0.7$\arcsec$ slit. Exposure times for each spectrum were 1200 seconds. For all of our sample, we stacked several spectra (from 2-4, depending on magnitude and sky conditions) in order to increment the S/N, obtaining final summed values ranging from about 50 to 90  (see Table \ref{param2}).

Data was reduced using the CarPy pipeline \footnote{\url{https://code.obs.carnegiescience.edu/mike}} \citep[][]{Kelson00,Kelson03} and included overscan subtraction, flat fielding, orders extraction, wavelength calibration and sky subtraction. Milky flat fields were taken with the aid of a diffuser to correct for the pixel to pixel variations.

We also conducted a search of high resolution spectroscopic archival data and found spectra of three stars useful for our study in the ESO archive (Phase 3) observed with the FLAMES-UVES spectrograph \citep{Dekker00,Pasquini03} (see Table \ref{param2}). 
UVES spectra of several other candidate NGC 1261 members exist but upon further analysis were found not to be genuine members based on Gaia proper motions and radial velocity.
The spectral range of the UVES spectra is between 4770 and 6830 \AA {},  with a resolution of $\sim$47000 (ESO program IDs 188.B-3002(M), 193.B-0936(N), 193.B-0936(O), 197.B-1074(G),193.D-0232(D)). This data has already been processed through the ESO pipeline and is ready for analysis. 
One of these stars turned out to be in common with our MIKE dataset. We reduced the two spectra independently.

We measured radial velocities using the FXCOR package from  IRAF and a synthetic spectrum as a template, which was generated using the mean of the atmospheric parameters of our stars, which are all found to be red giants. The mean heliocentric radial velocity value for our eight  targets
is 72.65$\pm$1.14 km$s^{-1}$, while the dispersion is 3.23  km$s^{-1}$ (see Table \ref{param2}).
The mean radial velocity is in very good agreement with the values in the literature: \citet{Baumgardt19}
found a value of 71.36$\pm$0.24 km$s^{-1}$ and \citet[2010 edition]{harris96} quotes a value of 68.2$\pm$4.6 km$s^{-1}$. Tables \ref{param1} and  \ref{param2} list the basic parameters of the selected stars: The J2000 coordinates, F275W, F336W and F438W magnitudes from \citet{Milone17}, heliocentric radial velocity (RV$_{H}$), Teff, log(g), micro-turbulent velocity (vt), metallicity ([Fe/H]),  number of iron lines measured for FeI and FeII, the final summed S/N of the spectra around 570 nm and 607 nm and the classification of the star as normal or anomalous using the above criterion.

\begin{figure*}
\centering

  \includegraphics[width=7.2in,height=3.3in]{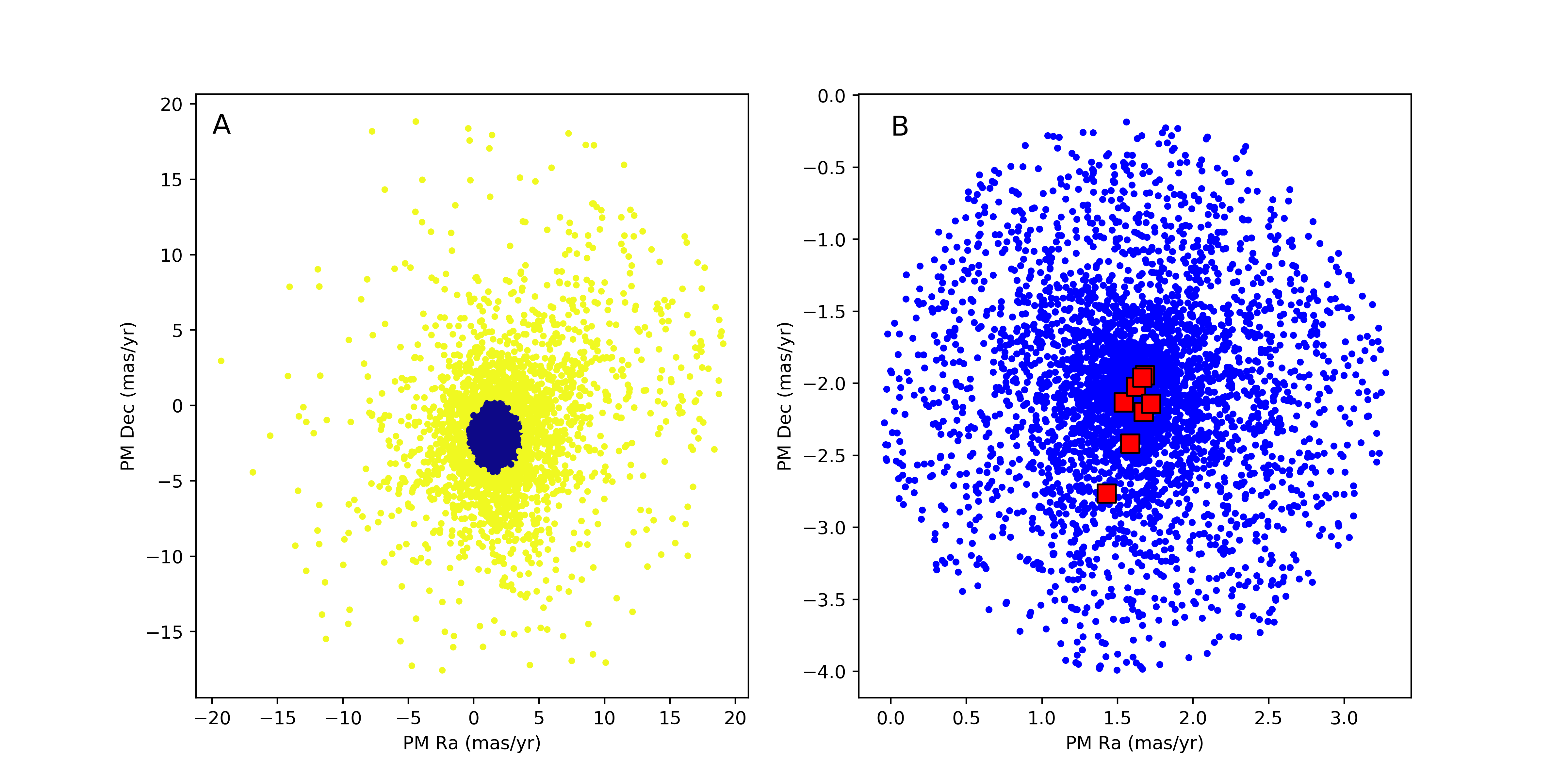}
  \caption{A and B show the PM distribution in the field of NGC 1261 from GAIA EDR3. A shows the PM distribution for all stars within the tidal radius of 5.05 arcmin from the center of NGC 1261 \citet [2010 edition]{harris96}. Yellow circles represent stars with lower membership probability and blue circles stars with higher membership probability. B is a zoom of the blue region where the red squares  are the seven  RGB stars analyzed in this study.}
  \label{gaiapm}
 \end{figure*}

We used the GAIA  EDR3  survey \citep{gaiacollaboration20}   to further analyze the membership of our targets using proper motion (PM). In figure \ref{gaiapm} we show
the position of our targets (filled red asterisks) in the PM plane. Clearly, all of our targets are high probability members of NGC 1261. We used a Gaussian Mixture Model (GMM) task from Python to estimate  membership, which  has been used in several studies   to determine membership through  PM space \citep{higuera2002,uribe2006,Cantat2019}. We also verified these results with previous PM studies \citep{Baumgardt18,raso20} and found  good agreement. Given these high PM membership probabilities, greater than 80\%, along with their common radial velocities as well as position in the various HST photometric diagrams and in the cluster, we consider all 8 stars as very likely members.




\begin{figure*}
\centering

  \includegraphics[width=7.0in,height=4in]{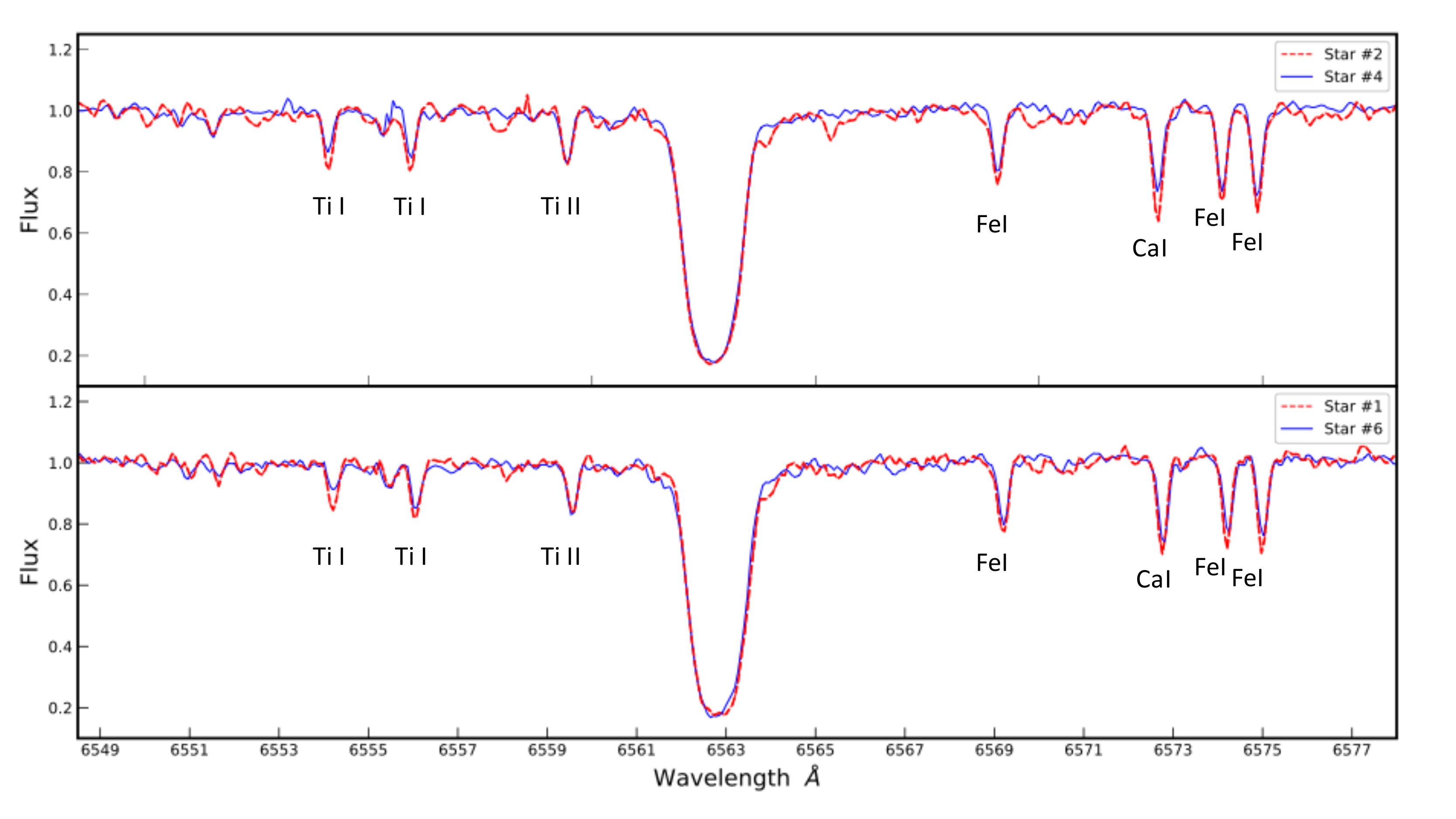}
  \caption{In the top of the figure comparison between star \#2 (anomalous ) and \#4 (normal) and   on the bottom of the figure comparison between star \#1 (anomalous ) and \#6 (normal) in the spectral region between 6550 to 6576{\AA} with H$\alpha$ in the center of the region, to which the fluxes were normalized. Both stars in each comparison  have very similar temperature but the anomalous stars has  stronger absorption in all metal lines, in particular FeI.}
  \label{halpha}
 \end{figure*}

\begin{figure}
\centering

  \includegraphics[width=3in,height=3in]{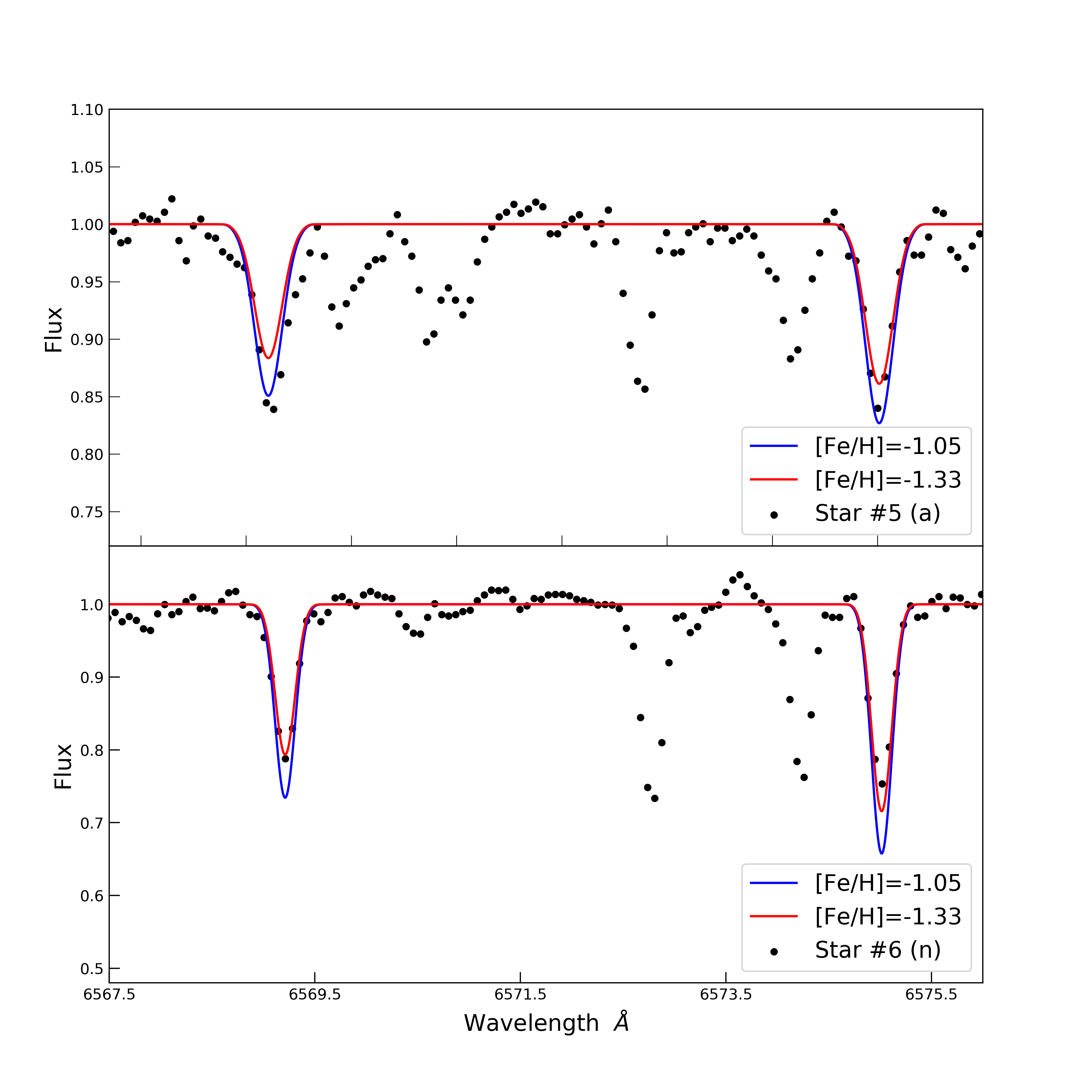}
  \caption{In the top,  comparison of  the observed spectrum of star \#5 with two synthetic spectrum (red and blue)  for two iron lines using the stellar parameters of star \#5 varying only the metallicity. In the bottom, comparison of the observed spectrum of star \#6 with two synthetic spectrum (red and blue)  for two iron lines using the stellar parameters of star \#6 varying only the metallicity. }
  \label{comp-synth}
 \end{figure}

\begin{table*}
\caption{Positions  and photometry from  \citet{Milone17}.}
\label {param1}
\centering
\small
\begin{tabular}{ c c c c c c c c c}
\hline
ID& GAIA ID  & RA  & DEC & F275W &  F336W & F438W & F814W &  F606W  \\
 &  &{\small{}(h:m:s)} & {\small{}($\,^{\circ}{\rm }$:$^{\prime}$:$^{\prime\prime}$ )}   & {\small{}(mag)} &{\small{}(mag)} & {\small{}(mag)} &{\small{}(mag)} &{\small{}(mag)}\\
\hline
1	& 4733794756152382336 & 03:12:09.77  & -55:12:49.92 & 19.670 & 17.397 &16.541 &14.293 &  15.127\\
2	& 4733794519928522496 & 03:12:18.24   &  -55:13:31.07 & 19.772& 17.397 &16.447&  14.085 &  14.940\\
3	& 4733794549993960832 &  03:12:13.13 & -55:13:57.11 &19.869& 17.115 &16.136 &13.649 &  14.552\\
4	& 4733794897888201216 & 03:12:13.48   & -55:11:47.76 &19.682 &17.142 &16.362& 14.050 &  14.909\\
5	& 4733794623008087936 & 03:12:20.63  & -55:12:27.27   &19.291& 17.603 &17.147& 14.953 &  15.697\\
6	& 4733794519931940864 & 03:12:18.28  & -55:13:44.31 &19.581&  17.221 & 16.566&14.377 & 15.211\\
7  &  4733794485571368192 & 03:12:16.51 &   -55:13:59.26 & 20.095 &17.134 & 15.894&13.152 & 14.126 \\
8  &4733794515634242048 & 03:12:19.70  & -55:13:22.60  & 20.050& 17.170& 15.920&13.206 & 14.176 \\

\hline
\end{tabular}
\end{table*}


\begin{table*}
\centering

\caption{Spectroscopic parameters of the observed stars}
\label{param2} \centering 
\begin{tabular}{ c c c c c c c c c c c c c   }
\hline 
{\small{}ID} &{\small{}RV$_{H}$ } & {\small{}T$_{eff}$} & {\small{}log(g)} & {\small{}{[}Fe/H{]} } & $v_{t}$ &  FeI & FeII & S/N  & S/N  & No.  & Class. & Inst. \tabularnewline

 & {\tiny{}(km $s^{-1}$)} & {\small{}{[}K{]} } &  & dex & {\small{}{[}km/s{]} } & & & (570\,nm) & (607\,nm) & Spectra& \tabularnewline
\hline

1   & 66.09 & 4558 & 1.49  & -1.31& 1.73 &113 & 12 & 60 & 60&3 & a & MIKE \tabularnewline
2   & 77.49 & 4456 & 1.39 & -1.18 & 1.51 &106 & 10 & 61 &71 &2 &  a& MIKE\tabularnewline
\hline 

3   & 75.92 & 4409 & 1.17  & -1.43 & 1.69 &106 & 10 & 73 & 90 &2 & n& MIKE\tabularnewline
3   & 75.04 & 4403 & 1.07  & -1.37 & 1.74 &60 & 5 & 44 & 48 &1 & n& UVES\tabularnewline

\hline 

4   & 69.55 & 4506& 1.26 & -1.32 & 1.69 &101 & 12 & 63 & 84&3  &n & MIKE\tabularnewline
5   & 69.91& 4858 & 2.05  & -1.05& 1.20 &65 & 9 &51 &53  &4 &a& MIKE\tabularnewline
6   & 70.49 & 4568 & 1.42  & -1.33& 1.62 &84 & 11 &81 & 67  &4&n& MIKE \tabularnewline

7   & 70.72 & 4264 & 1.03  & -1.40& 1.85 &58 & 6  &80 & 88  &1&n & UVES\tabularnewline
8   & 74.46 & 4279 & 1.00  & -1.39& 1.68 &61 & 6  &85 & 97  &1&n & UVES\tabularnewline

\hline 
\end{tabular} \\
Columns 7 and 8: Number of lines measured for FeI and FeII; Columns 9 and 10: S/N of the spectra in the spectral range mentioned.\\
Column 11:  Number of stacked spectra; Column 12: Classification of the stars, "n" for normal and "a" for anomalous.
\end{table*}


\begin{table}
\caption{Sensitivity of iron due to the uncertainties in atmospheric parameters for three stars representative in  NGC 1261}
\label {sens}
\centering
\begin{tabular}{ l c  c c  c  c c  c  c    }
\hline 

	ID   &  $\Delta T_{eff}  $ & $ \Delta log(g)$  & $\Delta v_{t}$	&  $\sigma_{S/N}$& $\sigma_{tot}$  \\		
      &  $42 K $ & $0.16$  & $ 0.10$& & \\	

	\hline

\#1     &	0.03   &	0.02   &	0.02  &	0.01   &	0.04 \\ 	
\#3     &	0.03   &	0.02   &	0.01  &	0.01   &	0.04 \\ 	
\#5     &	0.04   &	0.02   &	0.04  &	0.02   &	0.06 \\

\hline

\end{tabular}
\end{table}


\begin{table}
\caption{Parameters of the observed stars obtained from VHS photometry and equation 2}
\label{param_phot} \centering 
\begin{tabular}{ c c c c c c c c  c  c  }
\hline 
{\small{}ID} & {\small{}T$_{eff}$} & {\small{}log(g)} & {\small{}{[}Fe I/H{]} } & {\small{}{[}Fe II/H{]} }& $v_{t}$ & Class. \tabularnewline

  & {\small{}{[}K{]} } &  & dex & dex& {\small{}{[}km/s{]} } \tabularnewline
\hline

1   & 4583 &1.58 &  -1.27 & -1.26&1.53 &  a\tabularnewline  
2   & 4501 &1.45 &  -1.16 & -1.10 &1.56&a\tabularnewline
3   & 4418 &1.24 &  -1.40 &-1.36 & 1.61&n\tabularnewline
4   & 4522 &1.46 &  -1.29 &-1.24 &1.56&n\tabularnewline
5   & 4966 &1.88 &  -1.02 &-1.09& 1.45&a\tabularnewline
6   & 4622 &1.62 &  -1.26 &-1.27& 1.52&n\tabularnewline
7 & 4271 & 1.00 & -1.36&-1.38& 1.72 &n\tabularnewline
8 & 4352 & 0.81 & -1.34 & -1.39& 1.73 &n\tabularnewline

\hline 
\end{tabular} 

\end{table}

\section{Atmospheric Parameters and iron Abundances}

Atmospheric parameters were calculated using the same procedure described in several  previous studies \citep{munoz13,villanova13,mura17,munoz17,munoz18,munoz20} and the line list for the chemical analysis is the same described in   \citet{villanova11}.

The first set of stellar parameters were calculated directly from the spectra as follows. The atmospheric models were calculated using ATLAS9 (Kurucz 1970) and the stellar parameters  Teff, $v_t$, and log(g) were adjusted and new atmospheric models calculated iteratively in order to remove trends in excitation potential and equivalent width vs. abundance for Teff and $v_t$ respectively, and  we used  FeI and FeII   to satisfy the ionization equilibrium to derive log(g).  The [Fe/H] value of the model was changed at each iteration according to the output of the abundance analysis. The Local Thermodynamic Equilibrium (LTE) program MOOG \citet{sneden73} was used for the abundance analysis   until convergence was reached. The final derived values are presented in Table \ref{param2}.

We also carried out an internal error analysis  for the iron abundance, following the procedure described by \citet{marino08},  varying the stellar parameters ($T_{eff}$, log(g) and $v_t$) and redetermining the iron abundances of star \#1, which is representative of our sample. Parameters were varied by $\Delta$Teff = 42K, $\Delta$log(g)=0.16 and $\Delta v_t$=0.1 km/s, which are the  uncertainties in atmospheric parameters  for Teff, logg and vt respectively. These values were used to determine the sensitivity and the internal total error for iron. The amount of variation of the parameters was calculated using three stars representative of our sample (\#3,\#1, and \#5) with relatively low, intermediate and high effective temperature respectively. We calculated the error of the fit in the relation between abundances vs. EP  for all our sample, and then  obtained the average of this error which is the typical error of the fit. Then we fixed the other parameters (log(g), $v_t$ and [Fe/H]) and we varied the temperature in the three representative stars until the slope of the best fit in the abundance vs EP relation is equal to the corresponding mean error. The difference in the temperature is our estimation for $\Delta$Teff. An identical procedure was followed to obtain $\Delta v_t$  but in this case using the relation between abundance vs. EW. Since  we used the condition that the Fe abundance from FeI  has to be the same as that from FeII lines in order to obtain the gravity, the procedure  to obtain the error on gravity required  first obtaining  the  $\sigma$ for FeI and FeII, and afterwards we varied log(g) of the representative stars in order to satisfy the relation FeI-$\sigma_{FeI}$=FeII+$\sigma_{FeII}$. The error $\sigma$ was obtained dividing the rms scatter by the square root of the number of FeI and FeII lines.

The uncertainty on the EW ($\sigma_{S/N}$) was
calculated by dividing the rms scatter by the square root of the
number of the lines used to obtain the iron abundance.
In table \ref{sens}  we presented the sensitivity of iron due to the uncertainties in atmospheric parameters and uncertainties due to the errors in the EWs for the three representative stars (\#1,3, and 5), as well as the total. Also, in the same table we presented the total internal  error ($\sigma_{tot}$) for [Fe/H] as a result of the uncertainties in the atmospheric parameters and S/N which is  given by:\\

\begin{center}
$\sigma_{tot}=\sqrt{\Delta Teff^{2}+\Delta log(g)^{2}+\Delta v_{t}^{2}+\sigma^{2}_{S/N}}$\\
\end{center}

Finally, we take the  total internal error   from  star \#5 ($\sigma_{tot}$=0.06), in order to avoid  underestimating  the error (see Table \ref{sens}).

The second, independent set of stellar parameters (Table \ref{param_phot}) were calculated as follows. $T_{eff}$ was derived from the J-K color from the VISTA  Hemisphere Survey \citep[VHS]{McMahon19}  using the relation of Alonso et al. (1999) and the
reddening (E(B-V)=0.01) from \citet [2010 edition]{harris96}. Note that we did not use 2MASS colors as we found several of our stars in this rather dense GC to be substantially crowded in the large pixel scale of 2MASS (2$\arcsec$) and much cleaner in the much smaller VISTA pixels (0.34$\arcsec$). Surface gravities were obtained from the canonical equation:

\begin{center}
\begin{equation}
log(g/g_\odot)=log(M/M_\odot)+4log(Teff/T_\odot)-log(L/L_\odot)
\end{equation}
\end{center}

\noindent where the mass was assumed to be 0.8$M_\odot$, and the luminosity was obtained from the absolute magnitude $M_v$ assuming an apparent distance modulus of $(m-M)_V$ = 16.09 \citep [2010 edition]{harris96}. The bolometric correction (BC) was derived by
adopting the BC:$T_{eff}$ relation from  \citet{buzzoni10}. Finally, the micro-turbulent velocity $(v_t)$ was obtained from the relation of \citet{marino08}. We then carried out the abundance analysis using the same programs as for the spectroscopic determination.

Comparing the atmospheric parameters derived from these two methods, we find mean differences of $\Delta T_{eff}$=+43$\pm36$K, $\Delta log(g)$=+0.08$\pm$0.13, $\Delta v_t$=-0.04$\pm$0.16km/s, and $\Delta [Fe/H]$=+0.04$\pm$0.02 dex, in the sense photometric - spectroscopic method. The offset in temperatures is in the sense and of the order predicted by \citet{mucciarelli20}. The 
differences for all parameters is quite small and within the errors expected. In particular, the mean metallicity difference is only 0.04 dex, ranging from +0.02 to +0.07 dex. From here on, we will use the spectroscopic results unless otherwise noted.

Comparing our results based on spectroscopy for star \#3, observed with both MIKE and UVES, we found excellent agreement for all parameters. In particular, the Fe abundances vary by only 0.06 dex. In the end, we
preferred to keep the stellar parameters obtained from the MIKE spectra, due to the wider spectral range, which allowed us to detect 106 lines of FeI, in contrast with only 50 in UVES. Even more critical is the number of FeII lines, which are essential for the ionization equilibrium. In MIKE data we detected 10 lines of FeII but  only 5 lines in UVES. In addition, the S/N from the MIKE spectrum  is almost double that of the UVES spectrum.

\begin{table*}
\centering
\caption{Mean metallicity and  dispersion, mass, fraction of 1G and Type II stars and origin for Type II GCs.}
\label {typeII}
\begin{threeparttable}

\begin{tabular}{ c c c c c c c}
 \hline
 GCs & [Fe/H]  & $\sigma$ 
 & Mass\tnote{a} &   {\scriptsize $N_{1}/N_{Tot}$\tnote{e}} & {\scriptsize $N_{TypeII}/N_{Tot}$\tnote{e}} & Origin\tnote{f} \\
  &  dex & & M$_{\odot}$&\\
\hline

NGC~362\tnote{b} 	&-1.025$\pm$0.056 & 0.057 & 3.45x10$^{5}$&  0.279$\pm$0.015  &0.075$\pm$0.009 & GE \\ 

NGC~1261	&-1.27$\pm$0.05 & 0.133 & 1.67x10$^{5}$ &  0.359$\pm$0.016 &0.038$\pm$0.006& GE \\

NGC~1851\tnote{b}	& -1.033$\pm$0.077& 0.028 & 3.02x10$^{5}$ &   0.264$\pm$0.015 &  0.030$\pm$0.014 & GE \\

$\omega$ Cen\tnote{b}	& -1.511$\pm$0.077& 0.189 &3.55x10$^{6}$&   0.086$\pm$0.010  &0.640$\pm$0.018 & Seq\\

NGC~5286\tnote{c,g}	& -1.727$\pm$0.014&  0.103 & 4.01x10$^{5}$ &  0.342$\pm$0.015  &0.167$\pm$0.010 & GE\\

NGC~6388\tnote{c,h}	&  -0.428$\pm$0.008&  0.054 & 1.06x10$^{6}$ & 0.245$\pm$0.010    &0.299$\pm$0.016& Seq\\

M~22\tnote{b}	&  -1.524$\pm$0.092&  0.064 & 4.16x10$^{5}$&   0.274$\pm$0.020  &0.403$\pm$0.021& MD \\

NGC~6715\tnote{c,i}	&  -1.449$\pm$0.022&  0.183 & 1.41x10$^{6}$&  0.267$\pm$0.012  &0.046$\pm$0.011 & Sgr \\

NGC~6934\tnote{d}	& -1.26$\pm$0.07 & 0.122  & 1.17x10$^{5}$ &   0.326$\pm$0.020  &0.067$\pm$0.010& GE\\

NGC~7089\tnote{b}	&  -1.402$\pm$0.055&  0.042 & 5.82x10$^{5}$& 0.224$\pm$0.014   &0.043$\pm$0.006&GE\\

\hline
\end{tabular}
\begin{tablenotes}
{\scriptsize
\item a: Mass from \citet{Baumgardt18}; b: Metallicity and dispersion from \citet{meszaros20}; c: Metallicity and dispersion  from \citet{Bailin19}; d: Metallicity and dispersion   from \citet{marino18} using GRACES spectra; e: $N_{1}/N_{Tot}$ and  $N_{TypeII}/N_{Tot}$  from \citet{Milone17}; f: Origin from \citet{bajkova20}, specifically Gaia-Enceladus (GE), Sequoia (Seq), Sagittarius (Sgr) and Main Disk (MD); g:Metallicity and dispersion from \citet{Marino15} ; h: Metallicity and dispersion from \citet{Carrettea18,carretta09b}; i:Metallicity and dispersion from \citet{carretta10c} .  .
}
\end{tablenotes}
\end{threeparttable}

\end{table*}
\begin{figure}
\centering

  \includegraphics[width=3.8in,height=3in]{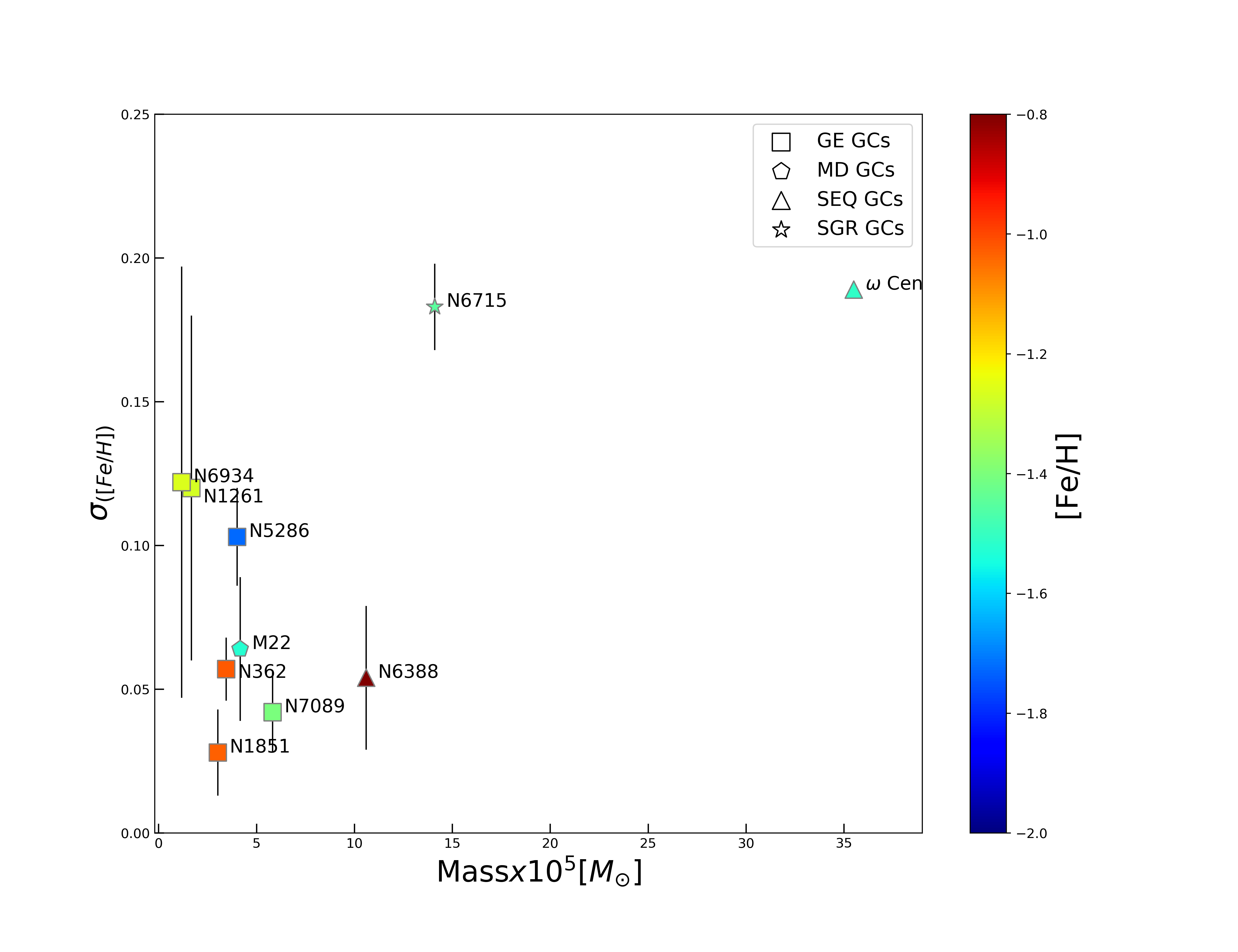}
  \caption{ $\sigma$([Fe/H]) vs Mass with color code by their metallicity     for all   the Type II GCs from Table \ref{typeII}. GCs origin is shown by the symbols in the legend from \citet{bajkova20}.}
  \label{mass_sigma}
 \end{figure}
\begin{figure}
\centering

  \includegraphics[width=3.3in,height=3.4in]{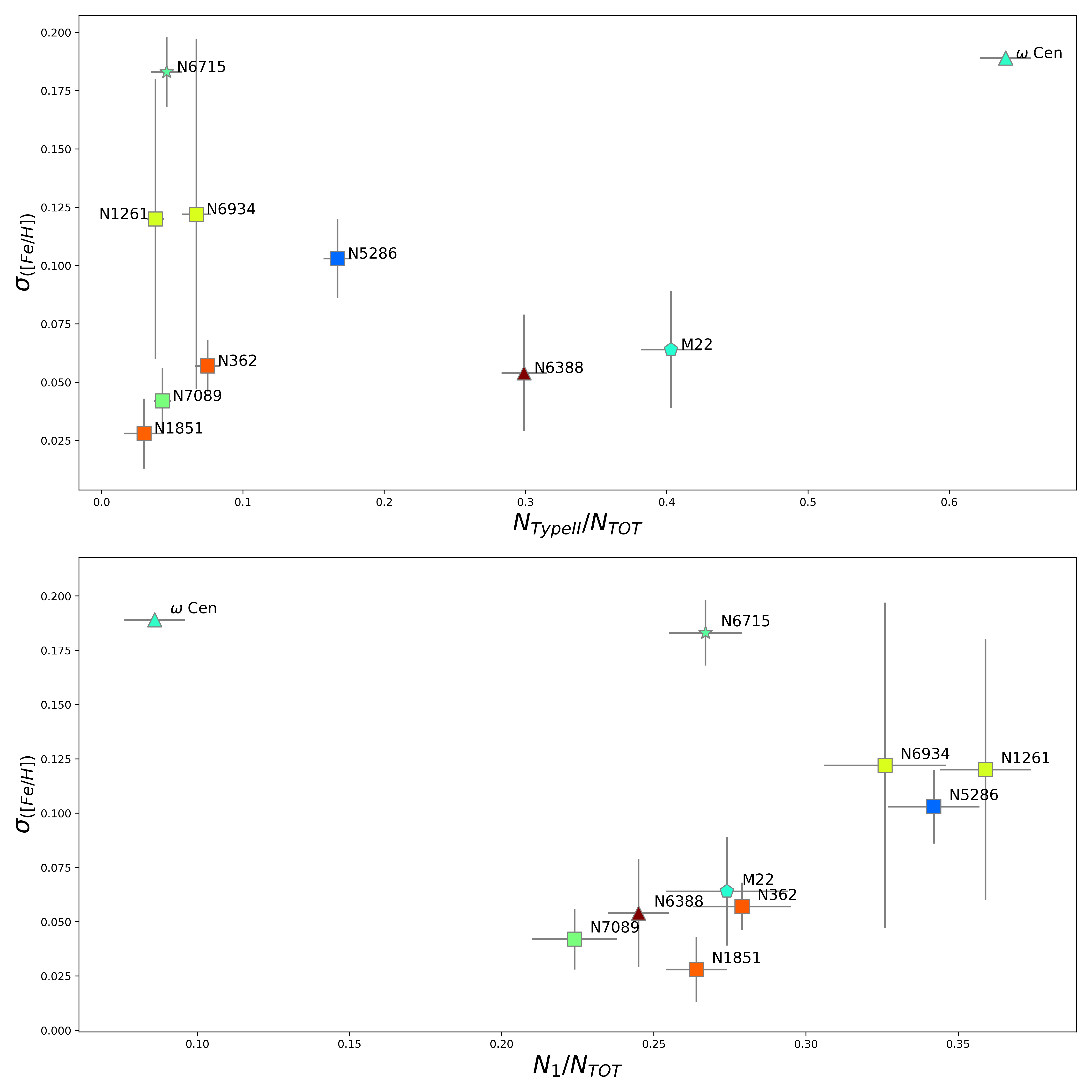}
  \caption{  $\sigma$([Fe/H]) vs. $N_{TypeII}/N_{Tot}$ (top) and  vs. $N_1/N_{Tot}$ (bottom)  for Type II GCs. The color code by metallicity and symbols  are the same as in Figure \ref{mass_sigma} }
  \label{Ntype_sigma}
 \end{figure}

\section {Results}

We focus our metallicity analysis of  NGC~1261 on any potential iron variation, in particular between normal and anomalous stars. 
We have studied eight upper RGB stars in NGC 1261, five belonging to the normal group and three to the anomalous group, using high resolution spectroscopy, and  measured the Fe abundance based on approximately 100 lines between FeI and FeII species in each star. It is worth noting the lack of previous studies of NGC 1261  with high resolution. To our knowledge, except for two  very recent articles \citep{sanna20,Koch21}, no other high resolution spectroscopic analysis has been published, despite the cluster's accessibility given its relative proximity, low reddening and substantial mass. Although our search did uncover spectra in the ESO UVES archive for several members, these had not been analyzed previously.

The Harris (2010) metallicity of -1.27 comes from an assortment of integrated photometric and spectroscopic indices, color-magnitude diagram analysis and low resolution Ca triplet spectra of 10 stars (Rutledge et al. 1997). The mean metallicity found in our study is <[Fe/H]>=-1.30$\pm$0.04, in good  agreement with the \citet [2010 edition]{harris96} value, but now much more robustly determined. Also, our results show good accord with the study presented by \citet{Koch21} using high resolution spectroscopy, who found a metallicity of  <[Fe/H]>=-1.26 dex with a scatter of 0.02 dex, but only for two stars.
 
In our  sample of eight  stars we find 
metallicities ranging from -1.05 to -1.43 for a total range of 0.38 dex, with a 
spread (standard deviation) of 0.119 dex. However, due to the small size of our sample, we prefer to calculate the spread ($\sigma$) using an alternative method to  verify the reliability of our measurements. This method was presented in the book "Introduction to Statistical Inference" \citep{Keeping}, which is useful for samples smaller than 20. Basically we multiply the range (R) of our sample by a factor $\kappa$, which depends on the size of the sample (N). Then we obtain the corrected sigma for our samples with the equation $\sigma=\kappa R$. This formula yields 0.133, slightly larger than the standard deviation. In order to detect an intrinsic spread, we must first account for the internal measurement errors. These were estimated above as 0.06 dex. Subtracting this in quadrature from the observed spread leaves an intrinsic spread of $\sim$0.119 dex. Thus, we do indeed detect a real variation in metallicity in NGC 1261 with high confidence.

More importantly, we are seeking to test the 
\citet{Milone17} hypothesis that there should be an intrinsic difference in metallicity between normal and anomalous stars. The most straightforward test is to directly compare the spectra of a normal and anomalous star with very similar derived temperatures. This is shown in Figure \ref{halpha}. The temperatures of the stars compared in the top of the figure differs by only 50K (only 21K photometrically), while their metallicities are offset by 0.14 dex (0.13 dex photometrically). The anomalous star shows significantly stronger FeI, CaI and TiI lines in the range between 6550 to 6576 \AA \ compared to the normal star, but the H$\alpha$ profiles  are almost identical, supporting the evidence that both stars have the same $T\rm _{eff}$ within a few tens of Kelvins. Similarly, the pair  of stars compared in the bottom pannel of the Figure \ref{halpha} (\#1 and \#6) differ by 10K and their metallicity differs only by  0.02 dex, but the position in the CMD and the Chromosome maps identify them clearly as an  anomalous and normal star respectively, and in the figure we can observe again  that the anomalous star (\#1)
shows significantly stronger FeI, CaI an TiI lines,  although their difference is less than the other pair. Additionally, in figure \ref{comp-synth}  we have  compared the stars  \#5 (anomalous) and star \#6 (normal) for two lines of Iron. In the  top of the panel we plotted the observed spectra for stars \#5 and over-plotted two synthetic models  using  the parameters of the stars \#5 varying only the metallicity between -1.05 (red line)  and -1.31 (blue line). In the bottom of the figure we made the same  for the star \#6, using the stellar parameters of this star. The figure  show the good agreement between the synthetic spectrum versus the observed spectrum with the correspondent metallicity. Finally, the figure \ref{halpha} and \ref{comp-synth} are the   most graphic examples of an intrinsic metallicity spread between the normal and anomalous stars - small but robustly detected.

Looking in detail at the spectroscopic results for the stars in each group, we find a mean metallicity of
<[Fe/H]>$_{n}$=-1.37$\pm$0.02 for the normal stars vs.
<[Fe/H]>$_{a}$=-1.18$\pm$0.09 for the anomalous stars. There is an 0.19 dex difference between the two groups, at a significance level of $\sim$2.4$\sigma$, with the anomalous stars being more metal-rich than their normal counterparts, as predicted by \citet{Milone17}.
We must stress that our sample is small but note that \citet{marino18} found a very similar effect in NGC 6934 using only 4 stars, 2 in each group. The most metal-poor of the anomalous stars has only a slightly higher metallicity than the most metal-rich normal star but the means are significantly different.

Figure \ref{CMD} marks the two groups in the HST CMD, where the blue stars are normal and red anomalous, based on their position in this diagram. We include Dartmouth \citep{dotter08} isochrones for metallicities of -1.15 and -1.36, $[\alpha/Fe]$=+0.4 and an age of 10.75 Gyr \citep{vandenberg13}. Clearly, these isochrones do a good job of reproducing the reddest and bluest RGBs, as well as the bright and faint SGBs, although they are slightly too blue and bright for the bluest RGB stars. This latter is an additional suggestion that the actual metallicity spread is likely somewhat less than 0.21 dex, consistent with the value of 0.18 dex we derive.
In the upper insert we see that the isochrones also agree well with the positions of our stars based on their atmospheric parameters, whose individual metallicites are shown in the lower insert.

The results based on the photometric atmospheric parameters are very similar. The
full range covers -1.02 to -1.40 dex, or 0.38 dex, with an intrinsic spread of 0.133 dex.  The normal stars have a mean  of <[Fe I/H]>$_{n}$=-1.33$\pm$0.03 and the anomalous stars a mean  of <[Fe I/H]>$_{a}$=-1.15$\pm$0.08, with a difference of 0.18 dex, significant at the $\sim$2.2$\sigma$ level.

The intrinsic metallicity spread is supported by the measurements which we made for Fe II (Table \ref{param_phot}). 
 The normal stars have a mean  of <[Fe II/H]>$_{n}$=-1.33$\pm$0.03 and the anomalous stars a mean  of <[Fe II/H]>$_{a}$=-1.15$\pm$0.06. This agreement is particularly important after   \citet{mucciarelli15} and \citet[]{Lardo16}, where the authors suggest that FeII lines are better spectral features in order to determine an intrinsic iron spread of a cluster, since  FeI lines can be affected by NLTE effects.
However, this effect is believed to be significant only for for metallicities below -1.50 dex \citep{mucciarelli20}.  Another possible explanation could be variability \citep{llancaqueo21}.

Thus, we find that there is an intrinsic metallicity spread in NGC 1261, as predicted by \citet{Milone17}. In addition, also as they predicted, the anomalous stars lying along the reddest RGB, which also lie to the red of the main locus of stars in the chromosome map, have higher metallicities than normal, bluer stars, by about 0.2 dex. This Type II GC now joins several other Type II GCs that have been well studied in hosting such a variation.

\section{Discussion and Conclusions}

Having verified the original hypothesis set forth by \citet{Milone17}, i.e. that there should be a real metallicity spread in the Type II GC NGC 1261 and in particular that anomalous stars should have a higher metallicity than normal stars, we turn to an investigation of Type II GCs in general to first update the latest knowledge regarding the status of genuine metallicity variations in them and to  explore any possible correlations between key parameters in an attempt to discern any explanation for their peculiar behavior, i.e. a genuine metallicity spread, and how such a spread might be related to the generic MP phenomenon. 

We begin by compiling a list of the Type II GCs and their relevant properties. These are presented in Table \ref{typeII}. This includes all ten of the Type II GCs classified by \citet{Milone17}. Note that there are several other GCs suspected or confirmed to possess real Fe abundance variations, including NGC 5824, NGC 6229, NGC 6273 and Terzan 5, which we do not include  here, since they were not observed as part of the HST UV Legacy Survey. For the metallicity measurements, we have relied on  \citet{meszaros20} when available. This is the most recent compendium of results based on the high resolution near-IR APOGEE spectrograph (Majewski et al. 2017). They analyzed all GC APOGEE data homogeneously, in particular with regard to any intrinsic dispersion. The number of stars in their Type II GCs ranges from 20 in M 22 to 775 in NGC 5139 and is typically 20-30. Their measure of dispersion ("scatter"), $\sigma_{int}$, is derived in the same manner as our value - subtracting $\sigma_{tot}$ in quadrature from $\sigma_{obs}$.

From this data, it is not clear that all of these Type II GCs actually have intrinsic spreads. If we look at their values for only Type I GCs from \citet{Milone17}, which are assumed not to possess such a spread, we find the mean scatter to be 0.10 dex for 21 clusters, with a large number of stars per cluster. If we assume this is the baseline and only clusters with scatters larger than this have real spreads, this rules out a real spread in NGC 362, NGC 1851, NGC 6388 and NGC 7089, almost half of our sample.  In particular, \citet{meszaros20} do not confirm the large metallicity spreads previously observed in M22 and NGC 1851 \citep{Marino09,carretta2010-1851}.  However, according to \citet{Milone17}, these GCs have anomalous star fractions of only 3-8$\%$ in the cases of NGC 362, NGC 1851, and NGC 7089, so previous studies not intentionally including such stars may well have missed them. However, a new study by Vargas et al. (2021, in preparation) designed to specifically investigate any metallicity variation in NGC 362 between normal and anomalous stars does not find any intrinsic dispersion.

For GCs not studied by \citet{meszaros20} we use the catalog of \citet{Bailin19}, who derive metallicity spread from high resolution spectra in the same way. Comparing his dispersion with that of \citet{meszaros20} for the seven Type II GCs in common, his values are even lower in NGC 362, NGC 1851, NGC 6388 and NGC 7089, further indicating that these Type II GCs may in fact NOT possess intrinsic metallicity spreads. The three GCs we use from his study are NGC 5286, with 62 stars,  NGC 6715 with 76 stars and NGC 6388 with 24 stars. The spread in the former is right at the above limit, while NGC 6715, the nuclear cluster of the Sagittarius dwarf spheroidal, does indeed have a significant spread.
Also, in the case of NGC 6388, we  preferred  to use the study  by  \citet{Bailin19} in our plots and in Table \ref{typeII}, because its  sample   is larger than the sample  by \citet{meszaros20} based on APOGEE data. In any case both studies show good agreement. \citet{meszaros20} found a mean metallcity  of -0.44 dex while  \citet{Bailin19} obtained [Fe/H]=-0.43 dex. On the other hand  \citet{Bailin19} quoted a dispersion of 0.05 dex, while \citet{meszaros20} obtained a slightly higher value of 0.07.

Finally, besides our study of NGC 1261 using 8 stars, the metallicity values for the other GC, NGC 6934, are from the high-resolution study of 4 stars by \citet{marino18}. Thus, the sample sizes are generally 10 or more with the exception of these latter two studies. Although the metallicity spread was derived in a homogeneous fashion, it is possible that these latter two studies may actually over-(or under-)estimate this value given their limited sample sizes.
We also list the GC mass from \citet{Baumgardt18}, the fraction of 1G and Type II stars from \citet{Milone17} and the estimated progenitor origin of the GC from \citet{bajkova20}.

We first note that the  observed  spread  in metallicity of 0.119 dex in NGC 1261  is comparable with that of other Type II GCs, especially with that of NGC 6934, which has a very similar mean metallicity and mass, and substantially larger than the 5 GCs noted above with dubious spreads. In fact, as stated in the Introduction, these are the two lowest mass Type II GCs but their metallicity spreads are actually quite large. Scenarios for an intrinsic metallicity spread within a GC generally invoke the need for a large mass in order to retain supernova ejecta and self-enrich \citep[e.g.][]{morgan89}. In 
Figure \ref{mass_sigma} we plot metallicity spread vs. mass, color coded by mean metallicity. We do not see any trend with metallicity, complementing the study of \citet{Bailin19}, and also confirm his finding that the most massive GCs, $\omega$ Cen and NGC 6715, have the largest spreads. However, we have to bear in mind that these two "GCs" are almost certainly the remnant nuclei of dwarf galaxies and therefore have a chemical evolution history distinct from normal GCs, although this may in fact also be the case for any other Type II GCs with genuine metallicity spreads \citep{Milone17}. 

Of the remaining Type II GCs, albeit only a small sample, we find evidence for an inverse trend, with metallicity spread increasing with decreasing mass, contrary to the standard scenario. This may in part be spurious, attributable to small number statistics and indeed possible overestimation of the metallicity spread in the two least massive GCs, which have the smallest sample of stars.
Also, note from the above discussion that the baseline for possessing a real intrinsic variation may well be 0.1 dex, meaning only the upper half of our sample truly have a variation. 

We note that no Type II GC exists with a mass below that of NGC 6934, $\sim 10^5M_\odot$, despite the large number of GCs studied by \citet{Milone17} below this value, strongly hinting that this may well be the minimum mass required by a GC to self-enrich in the heavy elements. This is quite close to the value suggested by \citet{morgan89}. It would indeed be strange for the metallicity spread to increase with decreasing mass and then suddenly be cutoff at a minimum mass.
Note that \citet{Bailin19} found a trend of larger metallicity dispersion with higher luminosity for GCs with $L>10^5L_\odot$, which also suggests a similar mass limit, given the typical GC M/L ratio around 2 \citep{baumgardt20}. However, \citet{meszaros20} found no correlation of the Fe abundance spread on either mass, age or absolute visual magnitude. 
Also note that the other GCs with confirmed or suspected metallicity spreads - NGC 5824, NGC 6229, NGC 6273 and Terzan 5 \citep{marino18}- all have masses of $3\times 10^5M_\odot$ or more.
Conversely, a number of GCs more massive than 
$10^5M_\odot$ studied by \citet{Milone17} are not classified as Type II GCs and are therefore not expected to have metallicity variations, and indeed those that have been studied in this regard do not exhibit them (e.g. \citet{meszaros20}). Thus, although there may be a minimum (present-day) mass to self-enrich, another factor must be involved as well.

Of course, a huge unknown in all of this is how much mass loss has occurred in a GC since its birth. The most relevant mass for retaining SNe is of course the initial mass of the cluster, or at least the mass at the moment of any SNe explosions. Massive core-collapse SNeII occur within some 10Myrs of formation, while lower mass SNeIa could occur after $\sim$ a Gyr. Both of these produce Fe. However, the latter also have the characteristic signature of lower $\alpha$-element yields, which are generally not seen in GCs. Therefore, any heavy-element self-enrichment in GCs likely involved only SNeII and therefore occurred very soon after the cluster's formation. Although there are a variety of stellar mass-loss mechanisms that operate on a GC during its life, which depend on a variety of parameters including internal factors such as concentration as well as external factors like the orbit, etc., these would have had very little time to operate before SNeII explosions and thus one might expect any mass loss before such explosions (after initial gas ejection) to have been small to negligible, but of course subsequent mass loss over the ensuing 10+ Gyrs could be very large. On the other hand, current scenarios explaining MP require invoking a very large mass loss, at least on the time scale of whatever mechanism causes MP occurred, probably not longer than $\sim 10^8 yr$  \citep{conroy2011}.  It is currently unclear how much mass has been lost by GCs and we are therefore left with the proxy of its current mass. However, note that \citet{Milone17} find that the fraction of 1G stars anticorrelates with current mass, indicating this parameter is at least related to MP and presumably also metallicity spreads. In the case of NGC 1261, an extended stellar halo has been recently found by \citet{raso20}, indicating the cluster was indeed more massive in the past and thus more likely to have self-enriched.
It is of course critical to try and estimate  the GC mass when it was formed and when the MP phenomenon occurred in order to constrain and improve models of GCs and MP formation. 

In Figure \ref{Ntype_sigma} we investigate any correlation between metallicity spread and the fraction of Type II stars, as well as the fraction of 1G stars in the cluster. No clear correlation is evident in the first plot. 
However, if we again discount the extreme outliers of a distinct nature, $\omega$ Cen and NGC 6715, we see a convincing hint of an increase in the metallicity spread with an increasing fraction of 1G stars. Again, the two highest spreads are those in the two clusters with the smallest samples, but the trend is intriguing and suggests a potential link between the MP phenomenon and metallicity variation. However, we remind the reader that the baseline for having an intrinsic spread may well eliminate the five GCs with the lowest $\sigma$ and therefore eliminate any apparent correlation. On the other hand, note that \citet{Milone17} do find a significant inverse  correlation between the fraction of 1G stars and cluster mass, which we would also find combining this figure and Figure \ref{mass_sigma} for only Type II GCs. 

In terms of origin,  all the Type II GCs have been accreted by the Milky Way from various progenitors, except for M22,
dominated by the Gaia-Enceladus accretion (see Table \ref{typeII}).  However, NGC 6388 was classified as Main Bulge GC by \citet{massari19}, therefore as an in-situ GC, in contrast to the classification proposed by \citet{bajkova20}. Although it is tempting to suggest that most Type II GCs are not native to the Galaxy, the fact is that due to reddening constraints, the HST UV Legacy Survey of GCs observed only a very small fraction of in-situ GCs, which are almost exclusively found in the bulge and/or disk, with their accompanying reddening problems. Therefore, with the current sample, it is not possible to make any strong assertions about the origin of Type II GCs, except for the observation that they come from a variety of progenitors. Of the four suspected/confirmed GCs with metallicity variations but not included in the UV Legacy Survey, one is a Main Bulge GC, one is from Gaia-Enceladus, one from Helmi99 streams and the remainder classified as Low Energy, adding to the variety of origins for these interesting objects.

We have presented here the first detailed iron abundance analysis for the Type II GC NGC 1261, based on high S/N, high-resolution spectra of eight  RGB members. The stars were carefully selected to include 3 pairs of normal and anomalous stars, based on the classification by \citet{Milone17} from their position in the chromosome map and UV CMD, observed by us with the MIKE spectrograph at LCO. In addition, we discovered spectra of 3 members in the ESO UVES database, one of which matched with a MIKE star. We confirm the prediction by \citet{Milone17} that this GC should be a member of the rare group of GCs with an intrinsic metallicity variation.
We find a range in [Fe/H]  for the entire sample from -1.05 to -1.43 dex with   an  intrinsic metallicity variation $\sigma_{int}$=0.119.  We verified these results based on a purely spectroscopic analysis with an alternative method based on VISTA  photometry to derive the atmospheric parameters.  Moreover, we analysed the metallicity separation between the two groups. We find an offset in mean metallicity of 0.19 dex, significant at the 2.4$\sigma$ level, in the sense that the anomalous stars have a higher metallicity, based on the spectroscopic technique, with very similar results from the photometric method. The analysis of only FeII lines also supports are finding. This result is again in accord with the prediction by \citet{Milone17}.

We then examined the ensemble of main parameters available for the 10 GCs classified as Type II by \citet{Milone17} to search for any trends amongst them. The latest compilations of metallicities in these GCs, based on high S/N, high resolution spectra of large samples, do not bear out the existence of intrinsic metallicity spreads in several of them. Thus,
we find strong indications from existing data that a substantial fraction of Type II GCs presumed to have metallicity spreads may in fact not possess them.
The data suggest that the minimum (current-day) mass of a GC with an intrinsic metallicity variation must exceed $10^5M_\odot$, although this is not a sufficient criterion.  
There is an intriguing suggestion that a larger metallicity spread is found in clusters with a larger fraction of 1G stars, hinting at a possible link between the MP phenomenon and metallicity variation, but this trend disappears if we eliminate GCs without a sufficient spread to claim a real variation.
Other than this, no strong trend emerges when looking for correlations between
metallicity, metallicity spread,  fraction of Type II stars and cluster mass and origin. Unfortunately, our knowledge of cluster mass is limited to its current value, while all of the interesting evolution leading to MP and metallicity variation, in which the actual cluster mass at that moment must have played a key role, occurred within the first Gyr or less of these primordial objects. 

According to the terminology presented by \citet{Milone17}, NGC~1261 would be classified as an Iron-II/TypeII GC, since it shows a significant variation in iron compared to other Galactic GCs. A refined classification is presented by \citet{marino18}, which refers to any variation in the s-process elements, as some Iron-II GCs also show a variation in s-process elements while others do not, such as NGC~6934 \citep{marino18}. Consequently a study of  other chemical species in NGC~1261, especially light elements, iron-peak and s-process elements, will be essential to further classify this intriguing object and help understand its formation and evolution in more detail. Such analysis of our spectra is currently in progress. In addition, of course, a larger sample of stars is always welcome. 
Also, a more detailed  chemical characterization of other Type II GCs will be crucial  to understand how abundances of particular elements, in addition to Fe, 
effect other parameters, such as their  position in the UV CMD and  the chromosome maps. In this respect, \citet{cummings14}  found that the CNO variations in  NGC 1851  played a important role.

We note here that it is essential to investigate detailed abundances in other Type II GCs not yet well studied, including NGC 6934 and NGC 362. \citet{meszaros20} finds no indication for intrinsic variation in this latter cluster from a large sample of APOGEE stars. However, they have not examined whether there is any metallicity offset between normal and anomalous stars. We have obtained MIKE spectra analogous to those reported on here for a sample of 6 pairs of normal and anomalous stars in this GC. We find (Vargas et al. 2021, in preparation) that there is no significant metallicity offset between the normal and anomalous stars, reinforcing the finding by \citet{meszaros20} that this is yet another Type II GC without a real metallicity spread. In addition, it is important to reassess any metallicity spread in the other Type II GCs that do not have strong indications for such a spread, at least from \citet{meszaros20} and \citet{Bailin19}. It is clear that large, homogeneous samples are crucial to robustly investigate internal cluster metallicity variations. In particular, it is of interest to search for any metallicity offsets between normal and anomalous stars based on the HST chromosome map and UV CMD. A plot like that of Figure \ref{mass_sigma} but with the mean difference in metallicity between normal and anomalous stars as abscissa would hopefully be revealing.

As we have noted, there are several other GCs suspected or confirmed to possess real Fe abundance variations, including NGC 
5824, NGC 6229, NGC 6273 and Terzan 5, which we do not include here as they were not observed as part of the HST UV Legacy Survey. It would be of great interest to study them in more detail and in particular classify them using the chromosome map and UV CMD. 
Such studies of more in-situ (bulge or disk) GCs would help to further constrain any role played by the origin of the cluster.

Finally, now that Gaia EDR3 is available, providing high confidence proper motions, as well as very reliable radial velocities from such projects as APOGEE, it would be very important to perform estimates of the total mass loss that a given GC has suffered over its lifetime, knowing its current location, age and orbit, using the most sophisticated modelling techniques available. This would help to clarify the role of the key parameter, mass, in leading to both MP and metallicity variation within a cluster.

\section*{Data availability}
The spectroscopic raw data analysed here was observed under the CNTAC Programm ID CN2018B-71, PI D. Geisler.
This work has made use of data from the European Space Agency (ESA) mission
{\it Gaia} (\url{https://www.cosmos.esa.int/gaia}), processed by the {\it Gaia}
Data Processing and Analysis Consortium (DPAC, \url{https://www.cosmos.esa.int/web/gaia/dpac/consortium}).

\section*{Acknowledgements}

The authors thank  support from Chilean Time Allocation Committee (CNTAC) through  program ID CN2018B-71.
SV gratefully acknowledges the support provided by Fondecyt reg. n. 1170518. 
C.M. thanks the support provided by  FONDECYT Postdoctorado No.3210144.
HF acknowledges financial support from Agencia Nacional de Investigacion y Desarrollo (ANID) grant 21181653.
D.G. and C.M. gratefully acknowledges support from the Chilean Centro de Excelencia en Astrof\'isica
y Tecnolog\'ias Afines (CATA) BASAL grant AFB-170002.
D.G. also acknowledges financial support from the Direcci\'on de Investigaci\'on y Desarrollo de
la Universidad de La Serena through the Programa de Incentivo a la Investigaci\'on de
Acad\'emicos (PIA-DIDULS).
The author would like to thank the referee for his valuable comments and suggestions.

\bibliographystyle{mnras}
\bibliography{biblio.bib}
\bsp	
\label{lastpage}
\end{document}